\documentclass[aps,prd,reprint,preprintnumbers,superscriptaddress,nofootinbib,floatfix]{revtex4-1}

\usepackage{amsmath,amssymb,amsfonts,dcolumn,color,graphicx,graphics,latexsym,placeins,epsfig}
\usepackage{epsfig}
\usepackage{bm}
\usepackage{slashed}
\usepackage{latexsym}
\usepackage{natbib}
\usepackage{url}
\usepackage{dcolumn}
\usepackage{color}
\usepackage{amsfonts,amssymb,amsmath}
\usepackage{graphicx,epsfig}
\usepackage{psfrag}
\usepackage{subfigure}
\usepackage{tabularx}
\usepackage{hyperref}
\hypersetup{colorlinks=true}
\usepackage{comment}

\newcommand{\be}{\begin{equation}}
\newcommand{\ee}{\end{equation}}
\newcommand{\ba}{\begin{eqnarray}}
\newcommand{\ea}{\end{eqnarray}}

\renewcommand{\inf}{\infty}

\begin{document}

\preprint{IPPP/26/26}

\title{Searching for axions with time resolved pulsar polarimetry}

\newcommand{\IPPP}{\affiliation{Institute for Particle Physics Phenomenology (IPPP), Department of Physics, Durham University, Durham DH1 3LE, United Kingdom}}

\author{Francesca Chadha-Day}
\email{francesca.chadha-day@durham.ac.uk}
\IPPP
\author{Tanmay Kumar Poddar}
\email{tanmay.k.poddar@durham.ac.uk}
\IPPP

\begin{abstract}
Pulsars possess strong dipole magnetic fields that can source axion fields through the axion-photon interaction. Pulsars may therefore be surrounded by axion field configurations oscillating with the pulsar's rotational period. These axions could be detected by observing their effect on the polarization of the pular's emission. In this paper, we use time resolved observations of the optical polarization of the Crab pulsar to place bounds on the axion-photon coupling, demonstrating the potential of time resolved pulsar birefringence in the search for axions.

\end{abstract}

\maketitle

\section{Introduction}
Pseudoscalar axions arise in extensions of the Standard Model (SM) that implement the Peccei-Quinn (PQ) mechanism to solve the strong CP (Charge conjugation-Parity) problem \cite{Peccei:1977hh,Weinberg:1977ma,Peccei:1977ur,Wilczek:1977pj}. The QCD (quantum chromodynamics) Lagrangian admits a CP-violating term which would induce a neutron electric dipole moment far larger than experimental limits unless the CP-violating parameter is unnaturally small \cite{Adler:1969gk,Bell:1969ts,Peccei:2006as,Baker:2006ts,Kim:2008hd}. In the PQ framework, a global chiral symmetry is spontaneously broken, giving rise to a massless Goldstone boson. Non-perturbative QCD effects explicitly break this symmetry through instantons, generating a periodic potential that dynamically relaxes the effective CP violating parameter to zero and endows the axion with a small mass. As a result, the axion mass is directly related to the PQ symmetry-breaking scale through QCD instanton effects \cite{GrillidiCortona:2015jxo}. In contrast, many extensions of high-energy theory, particularly string theory, generically predict a large number of ultralight pseudoscalar fields originating from the breaking of approximate global symmetries at very high energy scales \cite{Preskill:1982cy,Abbott:1982af,Dine:1982ah,Svrcek:2006yi,Marsh:2015xka}. For these axion-like particles (ALPs), the mass is generated by ultraviolet physics rather than QCD, and therefore their masses and couplings are largely independent, forming the so-called ``string axiverse" \cite{Arvanitaki:2009fg}.

Both the QCD axion and generic ALPs are well-motivated dark matter (DM) candidates \cite{Hu:2000ke,Duffy:2009ig,Hui:2016ltb,Chadha-Day:2021szb,OHare:2024nmr}. These particles can also couple to SM fields, and in particular their interactions with photons and fermions may lead to observable signatures \cite{Profumo:2017hqp}. Moreover, axions can mediate macroscopic forces between objects composed of ordinary matter \cite{Moody:1984ba}. In this work, we focus on axion couplings to the electromagnetic (EM) field, and study how such interactions modify photon emission from pulsars and magnetized stars, leading to potentially observable distortions in the photon spectrum and polarization.

A wide range of cosmological, astrophysical, and laboratory searches have already placed significant constraints on the axion-photon coupling over a broad region of axion parameter space, with a summary provided in \cite{AxionLimits}. Around a rotating, strongly magnetized compact star, the EM fields generate a non-vanishing $F_{\mu\nu}\tilde{F}^{\mu\nu}\propto\mathbf{E}\cdot\mathbf{B}$ density outside the stellar surface, where $F_{\mu\nu}$ denotes the EM field strength tensor, and $\mathbf{E}$ and $\mathbf{B}$ stand for the electric and magnetic fields of the magnetized star, which acts as a source for a pseudoscalar axion field. We compute the long-range axion profile produced by the pulsar EM fields and study the effect of the axion profile on propagating radiation.

As EM waves traverse this axion background, the polarization of light can rotate due to axion-photon interactions, leading to birefringence. The same interaction modifies the photon dispersion relation and may induce small frequency shifts or apparent redshift effects. Precision measurements of the photon spectrum can therefore be used to detect or constrain such couplings.

Broadband polarization observations of the Crab pulsar and nebula, spanning optical to hard X-ray and soft $\gamma$-ray energies including optical polarimetry with the Hubble Space Telescope \cite{Moran:2013cla}, X-ray measurements by the Imaging X-ray Polarimetry Explorer \cite{weisskopf22,Gonzalez-Caniulef:2024qsf}, and $\gamma$-ray observations from INTEGRAL satellite \cite{Dean:2008zz} provide a sensitive probe of photon propagation in a magnetized environment and are therefore well suited to constrain axion-photon mixing.

Several studies have used polarization observations of the Crab source to probe the axion-photon coupling. The POLARBEAR experiment, a cosmic microwave background (CMB) polarization telescope, monitored the polarization angle of the Crab at millimeter wavelengths over multiple observing seasons and searched for a sinusoidal time variation induced by an oscillating axion DM background. The absence of any detectable periodic modulation constrains the coupling to $g_{a\gamma\gamma}<2.16\times 10^{-12}~\mathrm{GeV}^{-1}\times (m_a/10^{-21}~\mathrm{eV})$ for axion masses in the range $9.9\times 10^{-23}~\mathrm{eV}\lesssim m_a\lesssim 7.7\times 10^{-19}~\mathrm{eV}$ \cite{POLARBEAR:2024vel}. Similar projected sensitivities based on pulsar polarimetry have also been discussed in \cite{Liu:2019brz}, considering the axion as the DM. 

In addition to pulsar observations, the polarization of the CMB can also be affected by an oscillating axion DM background, which induces a rotation of the polarization angle. Measurements of the CMB polarization with the South Pole Telescope (SPT) have been used to search for such time-dependent birefringence, leading to a constraint on the axion-photon coupling, $g_{a\gamma\gamma}\lesssim 1.18\times 10^{-12}~\mathrm{GeV}^{-1}\times (m_a/1.0\times10^{-21}~\mathrm{eV})$ \cite{SPT-3G:2022ods}. 

In this present work, we show that the axion need not be assumed to constitute DM but is instead generated as a long-range field sourced by the EM fields of the pulsar. The resulting rotation of the polarization angle of the emitted radiation is largely independent of photon frequency, allowing it to be distinguished from plasma-induced Faraday rotation. Similar considerations motivate \cite{Fan:2025ixw}, where relativistic axions produced instead via the axion-nucleon interaction are considered, leading to a signal that varies with the pulsar's rotation, as we will also find below.

The remainder of the paper is organized as follows. In Section.~\eqref{sec2} we derive the long-range axion field profile sourced by the pulsar's EM fields. In Section.~\eqref{sec3} we determine the modified photon dispersion relation in this axion background, compute the resulting frequency shift between emission and detection, and evaluate the axion-induced birefringence angle. In Section.~\eqref{sec4} we obtain constraints on the axion-photon coupling using polarization measurements of pulsar emission and compare them with existing limits. Finally, in Section.~\eqref{sec5} we summarize and discuss our results. 

We use natural units with $\hbar = c = 1$ throughout the paper unless stated otherwise.

\section{Axion Sourcing by Pulsar Electromagnetic Fields}\label{sec2}
The EM fields exterior to a neutron star (NS), modeled as a rotating, magnetized sphere with a misaligned dipole moment, were derived in \cite{deutsch}. While the exact vacuum solutions involve spherical Bessel functions of the third kind, the leading-order terms yield a considerably simpler and widely adopted description of the magnetospheric fields. Restricting to the lowest multipole order, the magnetic-field components can be written as

\begin{equation}
\begin{aligned}
B_r &= \frac{ B_0 R^3}{r^3}\left(\cos\alpha \cos\theta+\sin\alpha\sin\theta\cos(\phi-\Omega t)\right),\\
B_\theta &= \frac{B_0 R^3}{2r^3}\left(\cos\alpha\sin\theta-\sin\alpha\cos\theta\cos(\phi-\Omega t)\right),\\
B_\phi &= \frac{B_0 R^3}{2r^3}\sin\alpha\sin(\phi-\Omega t),
\end{aligned}
\label{eq:1}
\end{equation}

where $r$, $\theta$, $\phi$ are the spherical polar coordinates in the non-rotating frame of the observer, $B_0$ denotes the surface magnetic field of the pulsar, $R$ its radius, $\Omega$ the angular velocity, and $\alpha$ the misalignment angle between the rotation axis and the magnetic dipole moment. We impose the boundary conditions that the tangential component of the electric field is continuous at $r=R$, whereas the normal component may be discontinuous across the boundary to obtain the corresponding electric field components as

\begin{widetext}
\begin{equation}
\begin{aligned}
E_r &= -\frac{B_0\Omega R^5}{2r^4}\left[\cos\alpha\left(3\cos^2\theta-1\right)
+\frac{3}{2}\sin\alpha\cos(\phi-\Omega t)\sin2\theta\right],\\
E_\theta &= -\frac{B_0 R^3\Omega}{2r^2}\left[\frac{R^2}{r^2}\cos\alpha\sin2\theta
+\sin\alpha\left(1-\frac{R^2}{r^2}\cos2\theta\right)\cos(\phi-\Omega t)\right],\\
  E_\phi &= \frac{B_0 R^3\Omega}{2r^2}\sin\alpha\cos\theta\sin(\phi-\Omega t)
  \left(1-\frac{R^2}{r^2}\right),
  \end{aligned}
  \label{eq:2}
  \end{equation}
  \end{widetext}
  
where a nonzero misalignment angle $\alpha$ leads to a precession of the magnetic dipole with respect to the rotation axis of the pulsar, enabling the emission of pulsed radiation. In the aligned-rotator limit, $\alpha\to 0$, the magnetic and electric fields become
\begin{equation}
\begin{split}
\mathbf{B}(r,\theta)=\frac{B_0 R^3}{r^3}\Big(\cos\theta \hat{r}+\frac{\sin\theta}{2}\hat{\theta}\Big),\\
\mathbf{E}(r,\theta)=-\frac{B_0 \Omega R^5}{2r^4}\Big[(3\cos^2\theta-1)\hat{r}+\sin2\theta\hat{\theta}  \Big]. 
\label{eq:3}
\end{split}
\end{equation}

The time-independent magnetic and electric field configurations given in Eq.~\eqref{eq:3} can source a static axion field outside the star $(r\gtrsim R)$. For a massless axion, the leading contribution to the axion profile is dipolar and is approximately given by
\begin{equation}
a(r,\theta)\approx \frac{g_{a\gamma\gamma}B^2_0 R^5\Omega}{15r^2}\cos\theta,
\end{equation}
which is valid provided the axion mass satisfies $m_a\lesssim 1/R$. Such a dipolar axion configuration can in principle mediate an additional interaction if the pulsar resides in a binary system with another compact object, leading to a force that scales at least as $1/r^3$ and exhibits a characteristic angular dependence. However, this interaction is significantly weaker than the Newtonian gravitational force and therefore is unlikely to produce competitive constraints from binary dynamics measurements. The detailed derivation of the time-independent axion field profile is presented in Appendix~\ref{appendix}. 

To study the time-dependent profile of a pseudoscalar axion field sourced by the EM fields surrounding a pulsar, we consider the axion-photon interaction described by the Lagrangian
\begin{equation}
\mathcal{L} \supset
\frac{1}{2}\partial_\mu a\partial^\mu a
 -\frac{1}{2}m_a^2 a^2 -\frac{1}{4}F_{\mu\nu}F^{\mu\nu}
-\frac{1}{4}  g_{a\gamma\gamma} aF_{\mu\nu}\tilde F^{\mu\nu},
\label{axion1}
\end{equation}
where $a$ is the axion field with mass $m_a$, $F_{\mu\nu}$ is the EM field-strength tensor, and $g_{a\gamma\gamma}$ denotes the effective axion-photon coupling. Therefore, the equation of motion of the axion field becomes
\begin{equation}
  (\Box + m_a^2)a = - g_{a\gamma\gamma} \mathbf{E}\cdot\mathbf{B},
  \label{axion2}
\end{equation} 

where $F_{\mu\nu}\tilde F^{\mu\nu}=4\mathbf{E}\cdot\mathbf{B}$, with $\mathbf{B}$ and $\mathbf{E}$ denote the magnetic (Eq.~\eqref{eq:1}) and electric (Eq.~\eqref{eq:2}) fields of the rotating magnetized NS, respectively. Therefore, the axion-induced charge density $\rho_a=g_{a\gamma\gamma}\mathbf{E}\cdot\mathbf{B}$ acts as a source in the axion equation of motion
\begin{equation}
(\Box + m_a^2)a =g_{a\gamma\gamma}\frac{B^2_0R^6\Omega}{8r^5}\sin2\alpha\sin\theta\cos(\phi-\Omega t),    
\label{axion3}
\end{equation}
where we use Eqs.~\eqref{eq:1} and \eqref{eq:2} for the EM fields of the pulsar. We neglect time-independent contributions, as they do not lead to axion radiation, and terms proportional to $\Omega r$, which are suppressed in the regime $\Omega R\ll 1$ and yield subdominant contributions compared to the time-dependent source term on the right-hand side of Eq.~\eqref{axion3}. In the limit $\alpha\to 0$, the axionic source charge density becomes zero and there would be no axion radiation. The axionic source charge density is dipolar in nature and its time-oscillating behavior results in time-dependent axion field profile. 

In the following, we obtain the time-dependent axion field profile in the far zone ($r >> R$). As we will consider emission from the edge of the pulsar light cylinder, this approximation is reasonable (as previously considered in \cite{Khelashvili:2024sup}). A more detailed discussion of the axion profile around pulsars including the interior and near field zones can be found in \cite{Garbrecht:2018akc}.

We start from Eq. \eqref{axion3} and decompose the axion field into a monochromatic mode at the rotation frequency $\Omega$ as
\begin{equation}
a(t,\mathbf{r})=\Re\left[a_\Omega(\mathbf{r})e^{-i\Omega t}\right].
\label{axion5}
\end{equation}
The mode function $a_\Omega$ satisfies
\begin{equation}
(\nabla^2+k^2)a_\Omega(\mathbf{r})=-\rho_\Omega(\mathbf{r}),
\qquad
k\equiv\sqrt{\Omega^2-m_a^2},
\label{axion6}
\end{equation}
where $\rho_\Omega$ denotes the spatial part of the oscillating source.
Using $\cos(\phi-\Omega t)=\Re[e^{i\phi}e^{-i\Omega t}]$, the complex source can be written as
\begin{equation}
\rho_\Omega(\mathbf{r})
=\mathcal{S}_0\frac{\sin\theta}{r^{5}}e^{i\phi},
\qquad
\mathcal{S}_0
=g_{a\gamma\gamma}\frac{B_0^{2}R^{6}\Omega}{8}\sin2\alpha .
\label{axion7}
\end{equation}

It is convenient to express the source in terms of the spherical harmonics $Y_{l,m}$ as
\begin{equation}
\rho_\Omega(\mathbf{r})
=\frac{\mathcal{C}Y_{1,1}(\theta,\phi)}{r^{5}},
Y_{1,1}=-\sqrt{\frac{3}{8\pi}}\sin\theta e^{i\phi},
\mathcal{C}=-\mathcal{S}_0\sqrt{\frac{8\pi}{3}} .
\label{axion8}
\end{equation}
Therefore, the axion-induced source is proportional to $(l,m)=(1,1)$ dipole component.

We can write the multipole coefficients at frequency $\omega=\Omega$ as
\begin{equation}
Q_{l,m}
=\int d^{3}r^\prime\rho_\Omega(\mathbf{r}^\prime)
{r^\prime}^{\ell}Y_{l,m}^{*}(\hat r^\prime),
\label{axion9}
\end{equation}
which for our axion case takes the form $Q_{1,1}=\mathcal{C}/R$.

In the far zone, the outgoing solution of axion field from Eq. \eqref{axion6} can be written as
\begin{equation}
a_\Omega(\mathbf{r})
=\frac{e^{ikr}}{r}\sum_{\ell,m}
i^\ell\frac{k^\ell}{(2\ell+1)!!}
Y_{l,m}(\hat n)Q_{l,m}.
\label{axion10}
\end{equation}

The dipole contribution $(l=1)$ to the axion field profile yields
\begin{equation}
a_\Omega(\mathbf{r})
=\frac{ik\mathcal{S}_0}{3R}
\frac{e^{ikr}}{r}
\sin\theta e^{i\phi} ,
\label{axion11}
\end{equation}
where
\begin{equation}
Q_{1,1}Y_{1,1}
=\frac{\mathcal{S}_0}{R}\sin\theta e^{i\phi} .
\end{equation}

Therefore, the physical axion field profile becomes
\begin{align}
a(t,\mathbf{r})
&=\Re\left[a_\Omega(\mathbf{r})e^{-i\Omega t}\right]\nonumber\\
&=-\frac{g_{a\gamma\gamma} B_0^{2} R^{5}k\Omega}{24r}
\sin\theta\sin2\alpha
\sin(\phi-\Omega t+kr),
\label{axion12}
\end{align}
which is valid in the far zone for axion mass $m_a < \Omega$. This is the explicit time-dependent axion field profile, which has a dipolar behavior with an angular pattern $\sin\theta\sin(\phi-\Omega t+kr)$ and the amplitude of the wave scales as $g_{a\gamma\gamma}B^2_0 R^5\Omega k/r$. The back reaction of the axion field on the pulsar's magnetic field is 2nd order in $g_{a \gamma \gamma}$ and so may be neglected.

\section{Photon Propagation and Axion-Induced Birefringence}\label{sec3}
The photon propagation through the axion field background of Eq.~\eqref{axion12} modifies Maxwell's EM equations.
Starting from the axion-photon Lagrangian in Eq.~\eqref{axion1}, one obtains
\begin{equation}
\begin{aligned}
\nabla\cdot\mathbf{E} &= -g_{a\gamma\gamma}\nabla a\cdot\mathbf{B}, \\
\nabla\times\mathbf{B} &= \frac{\partial\mathbf{E}}{\partial t}
+ g_{a\gamma\gamma}\left(\nabla a\times\mathbf{E}
+ \mathbf{B}\frac{\partial a}{\partial t}\right),\\
  \nabla\cdot\mathbf{B} &= 0, \\
  \nabla\times\mathbf{E} &= -\frac{\partial\mathbf{B}}{\partial t},
  \end{aligned}
  \label{axion_Maxwell}
  \end{equation}
in absence of background source plasma and current densities. The axion-induced modified Maxwell's equations yield wave equations for the  $\mathbf{B}$ and $\mathbf{E}$ fields of the photon as
\begin{equation}
\Box \mathbf{B}=g_{a\gamma\gamma}\nabla\times[\dot{a}\mathbf{B}+\nabla a\times \mathbf{E}],   
\label{wq1}
\end{equation}
and
\begin{equation}
\Box \mathbf{E} +\nabla(\nabla\cdot \mathbf{E})=-g_{a\gamma\gamma}\frac{d}{dt}[\dot{a}\mathbf{B}+\nabla a\times \mathbf{E}].
\label{wq2}
\end{equation}

To study the photon propagation, following the procedure employed in \cite{Blas:2019qqp}, we employ the geometric-optics (Eikonal) approximation \cite{PhysRev.126.1899}, valid when the photon wavelength is much smaller than the characteristic axion-field variation scale; equivalently, $\partial_\mu \partial_{\nu} a/\partial_{\rho} a\ll \partial_\mu \mathbf{B}/\mathbf{B}, \partial_\mu \mathbf{E}/\mathbf{E}$. In this regime, the EM fields can be written as
\begin{equation}
\mathbf{B}(x,t)=\bm{\mathcal{B}}e^{iS(x,t)},
\qquad
\mathbf{E}(x,t)=\bm{\mathcal{E}}e^{iS(x,t)},
\label{wq3}
\end{equation}
where $\bm{\mathcal{B}}$, and $\bm{\mathcal{E}}$ are the magnitudes of magnetic and electric fields of the radiation, respectively, and $S(x,t)$ denotes the rapidly varying phase, with
\begin{equation}
\omega=-\frac{\partial S}{\partial t},
\qquad
\bm{\kappa}=\nabla S .
\end{equation}
Neglecting derivatives of $\omega$, $\bm{\kappa}$, $\bm{\mathcal{B}}$, and $\bm{\mathcal{E}}$ whose variations are small, and discarding second derivatives of $a$, the wave equations for $\mathbf{B}$ and $\mathbf{E}$ reduce to
\begin{equation}
\Box \mathbf{B}-g_{a\gamma\gamma}[\dot{a}\nabla\times \mathbf{B}+\nabla a(\nabla\cdot \mathbf{E})-(\nabla a\cdot \nabla)\mathbf{E}]=0, 
\label{wq4}
\end{equation}
and
\begin{equation}
\Box \mathbf{E}-g_{a\gamma\gamma}[\nabla \mathbf{B}\cdot \nabla a-\dot{a}\dot{\mathbf{B}}-\nabla a\times \mathbf{\dot{E}}]=0. 
\label{wq5}
\end{equation}

Combining Eqs. \eqref{wq4} and \eqref{wq5}, we obtain
\begin{equation}
\mathbf{M}(\omega,\bm{\kappa})\cdot (\mathbf{E},\mathbf{B})^{T}=0,
\label{wq6}
\end{equation}
where $\mathbf{M}$ is the matrix constructed from the modified Maxwell's equations. The photon dispersion relation follows from the condition that the eigenvalues of $\mathbf{M}$ vanish. Diagonalizing the matrix, results
\begin{equation}
\lambda_{\pm}=\omega^{2}-|\bm{\kappa}|^{2}\pm g_{a\gamma\gamma}(\kappa\cdot\partial a),
\label{wq7}
\end{equation}
where $\kappa^\mu=(\omega,\bm{\kappa})$, $\partial_\mu a=(\dot{a},\nabla a)$, and
$\kappa\cdot\partial a=\kappa^\mu\partial_\mu a=\omega\dot{a}+\bm{\kappa}\cdot\nabla a$.
The two signs in Eq.~\eqref{wq7} correspond to right- and left-handed circular polarizations.

Using the standard Hamiltonian equations for photon trajectories, we write

\begin{equation}
\begin{aligned}
\frac{d\mathbf{x}}{dt}&=-\frac{\partial\lambda/\partial\bm{\kappa}}{\partial\lambda/\partial\omega}
=\frac{\bm{\kappa}\mp(g_{a\gamma\gamma}\nabla a/2)}{\omega\pm (g_{a\gamma\gamma}\dot{a}/2)},\\
\frac{d\bm{\kappa}}{dt}&=\frac{\partial\lambda/\partial\mathbf{x}}{\partial\lambda/\partial\omega}
=\pm g_{a\gamma\gamma}\frac{\omega\nabla\dot{a}+(\bm{\kappa}\cdot\nabla)\nabla a}{2\omega\pm g_{a\gamma\gamma}\dot{a}},\\
\frac{d\omega}{dt}&=-\frac{\partial\lambda/\partial t}{\partial\lambda/\partial\omega}
=\mp g_{a\gamma\gamma}\frac{\omega\ddot{a}+(\bm{\kappa}\cdot\nabla)\dot{a}}{2\omega\pm g_{a\gamma\gamma}\dot{a}}.
\label{wq8}
\end{aligned}
\end{equation}

Integrating the above equations between the emission point $(x_e,t_e)$ and the detection point $(x_d,t_d)$ yield the net variations in photon frequency and wave-number as
\begin{equation}
\begin{aligned}
\Delta\omega &= \mp\frac{g_{a\gamma\gamma}}{2}\big[\dot{a}(x_d,t_d)-\dot{a}(x_e,t_e)\big],\\
\Delta\bm{\kappa} &= \pm\frac{g_{a\gamma\gamma}}{2}\big[\nabla a(x_d,t_d)-\nabla a(x_e,t_e)\big],
\end{aligned}
\label{wq9}
\end{equation}

respectively. Since the axion field falls off as $1/r$ in the far zone, the contributions of frequency and wave-number variations at the detector is negligible compared to their contributions at the source. Thus the fractional frequency shift of pulsar light due to the axion background can be written as
\begin{widetext}
\begin{equation}
\frac{\Delta\omega}{\omega}\simeq
\frac{g_{a\gamma\gamma}^{2}B_0^{2}R^{5}\Omega^{3}}{48\omega r_e}
\left(1-\frac{m_a^{2}}{\Omega^{2}}\right)^{1/2}
\sin 2\alpha \sin \theta_e \cos(\phi_e-\Omega t_e+kr_e).
\label{main1}
\end{equation}
\end{widetext}
Using the Crab pulsar as a benchmark, with surface magnetic field $B_0=8.5\times 10^{12}~\mathrm{G}$ \cite{Khelashvili:2024sup}, radius $R=14~\mathrm{km}$ \cite{Khelashvili:2024sup}, spin period $P=2\pi/\Omega=33~\mathrm{ms}$ \cite{Lyne:2014qqa}, and the emission point at the light cylinder radius $r_e=R_L=1/\Omega$, we estimate the sensitivity to the axion-photon coupling from precision measurements of photon frequency shifts as

\begin{widetext}
\begin{equation}
\begin{split}
\frac{\Delta\omega}{\omega}\simeq 10^{-12}\Big(\frac{g_{a\gamma\gamma}}{10^{-12}~\mathrm{GeV}^{-1}}\Big)^2 \Big(\frac{B_0}{8.5\times 10^{12}~\mathrm{G}}\Big)^{2}\Big(\frac{R}{14~\mathrm{km}}\Big)^4\Big(\frac{\Omega}{(2\pi/33~\mathrm{ms})}\Big)^3\Big(\frac{1~\mathrm{GHz}}{\omega}\Big).
\end{split}
\end{equation}
\end{widetext}

Thus, the axion-induced frequency shift is too small to be observed. We therefore turn to study the birefringence of the pulsar light due to propagation through an axion background. Let the photon follows a trajectory
\begin{equation}
\mathbf{x}=\mathbf{x}(l),~~~t=t(l),    
\label{bire1}
\end{equation}
where $l$ is the path length or the affine parameter along the ray and $\hat{n}$ is the unit vector along the propagation direction, so that $\tfrac{d\mathbf{x}}{dl}=\hat{\mathbf{n}}$.
For a photon in flat space, we choose $l$ such that $\tfrac{dt}{dl}\simeq 1$.

Therefore, we can write, 
\begin{equation}
\frac{da}{dl}=\frac{\partial a}{\partial t}\frac{dt}{dl}+\nabla a\cdot \frac{d\mathbf{x}}{dl}=\dot{a}+\hat{\mathbf{n}}\cdot \nabla a.   
\label{bire4}
\end{equation}
Setting $\lambda_\pm=0$ in Eq. \eqref{wq7} and for radial propagation, with $\nabla a\parallel\bm{\kappa}$, we obtain
\begin{equation}
\kappa_r=\omega\pm\frac{g_{a\gamma\gamma}}{2}(\dot{a}+\partial_r a).
\label{wq10}
\end{equation}

Therefore, the phase shift between the left and right circularly polarized modes can be written as
\begin{equation}
\Delta \phi=\int _{\mathrm{path}}(k_r^{+}-k_r^{-})dr=\int _{l_{e}}^{l_{d}} g_{a\gamma\gamma}\frac{da}{dl}dl=g_{a\gamma\gamma}(a_o-a_e)
\label{bire5}
\end{equation}
where $a_o(a_e)$ denotes the axion field values at the observer (emission) point. Since, the axion field falls off as $1/r$, the field value at the emission point is much larger than the detection point.

Therefore, the birefringent angle, which is the half of the phase shift $\Delta\phi$, is obtained as
\begin{widetext}
\begin{equation}
\Delta\psi= \frac{g^2_{a\gamma\gamma}B^2_0R^5\Omega^2}{48 r_e} \Big(1-\frac{m^2_a}{\Omega^2}\Big)^{1/2}\sin 2 \alpha \sin \theta_e \sin(\phi_e-\Omega t_e+kr_e),
\label{bire6}
\end{equation}
\end{widetext}
where $(r_e,\theta_e,\phi_e)$ are the coordinates of the emission point. In practice, we view the pulsar as a point source, and so must integrate the birefringence signal over the entire hemisphere of the pulsar facing the Earth, as discussed in section \ref{sec4}.

Axion-photon interactions thereby induce an {\it oscillatory} birefringence signal in pulsar radiation, leading to a rotation of the polarization plane that is independent of the photon frequency and oscillates at the pulsar's rotation frequency. This birefringent angle can be directly probed through precision polarimetric measurements of pulsar emission. The effect operates for axion masses $m_a < \Omega$, set by the stellar rotation frequency, and does not require the axion to constitute the DM.

\section{Constraints on axion-photon coupling from Crab pulsar polarization}\label{sec4}

We consider the Crab pulsar as a benchmark source for probing the axion-photon coupling. It constitutes an excellent testbed owing to the availability of high-precision polarization measurements \cite{Slowikowska:2009rm,Moran:2013cla}. The optical linear polarization of the Crab pulsar has been measured with high time resolution in \cite {Slowikowska:2009rm} and is well described by the combination of an unpulsed component of fixed polarization angle and a pulsed component. The pulsed component displays sharp features at the main pulse and interpulse with a qualitatively different time dependence than the axion contribution given in Eq.~\eqref{bire6}. The unpulsed component is constant over the pulsar's rotation and is believed to be associated with the continuous direct current emission, whose source is unknown but is generally considered to originate in the magnetosphere or pulsar wind zone. We may therefore place bounds on the axion-photon coupling based on the non-observation of linear polarization angle modulation in the unpulsed component. 

\begin{figure}[ht]
    \centering
\includegraphics[width=\linewidth,keepaspectratio]{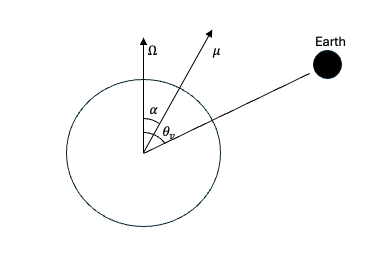}
    \caption{Schematic geometry of the pulsar as seen from Earth. The rotation axis of the pulsar is denoted by ${\bf \Omega}$, the magnetic dipole moment by ${\bm \mu}$ and $\theta_v$ denotes the viewing angle, defined as the angle between the pulsar rotation axis and the observer’s line of sight.}
    \label{angles}
\end{figure}

To illustrate this effect, we will assume that the unpulsed emission occurs uniformly at the radius of the light cylinder $R_L =1/\Omega$. To obtain the observed axion induced birefringence, we must integrate the Stokes parameters $(I,Q,U,V)$ including the birefringence angle in Eq.~\eqref{bire6} over the hemisphere of the pulsar facing the Earth. Assuming the linear polarizatation degree and angle, in the absence of axion effects, are uniform over the sphere at radius $R_L$, the Stokes vector angular density is

\begin{equation}
    d{\bf S}(\theta, \phi, \Omega t) = \frac{1}{2 \pi}\begin{pmatrix} I \\ Q \\ U \\ V \end{pmatrix} d\Omega 
    = \frac{1}{2 \pi}\begin{pmatrix} I \\ I p \cos (2\psi + 2\Delta \psi (\theta, \phi, \Omega t) \\ I p \sin (2\psi + 2\Delta \psi (\theta, \phi, \Omega t)) \\ 0 \end{pmatrix} d\Omega,
\end{equation}

where $p$ is the linear polarization degree in the absence of axions, $\psi$ is the linear polarization angle in the absence of axions, $d \Omega$ is the differential area element on the unit sphere and we have neglected any intrinsic circular polarization as this is not relevant to the effect considered here $(V=0)$. To obtain the linear polarization angle (depending on $U$ and $Q$) and degree (depending on $U$, $Q$, and $I$, where $I$ measures the intensity) in the presence of an axion field sourced by the pulsar's magnetic field, we must integrate the Stokes parameter density over the hemisphere of the pulsar facing the Earth:

\begin{equation}
{\bf S}_{\rm tot}(\Omega t) = \int_{\phi = 0}^{\frac{2 \pi}{\sin(\theta_v) + 1}} \int_{\theta = 0}^{\theta = \theta_v + \pi/2} d{\bf S},
\label{totalStokes}
\end{equation}
where $\theta_v$ is the viewing angle of the pulsar with respect to its rotation axis, defined to be in the range $0 \leq \theta_v \leq \pi/2$. The limits on the integral over $\phi$ are derived by considering the surface area of the hemisphere. The geometry is shown in FIG.~\ref{angles}

The observed linear polarization degree $p_{\rm lin} = \sqrt{Q^2+U^2}/I$ and angle $2\psi = {\rm tan}^{-1}(U/Q)$ can then be calculated from ${\bf S}_{\rm tot}$.

When $\theta_v = 0$, we find that there is no axion-induced change to the observed linear polarization angle, as birefringence contributions from different $\phi_e$ exactly cancel out for all $t_e$. However, for other viewing angles the full range of $\phi$ is not in view and so the sinusoidal oscillations in linear polarization degree and angle may be observed.

\begin{figure}[ht]
    \centering
\includegraphics[width=\linewidth,keepaspectratio]{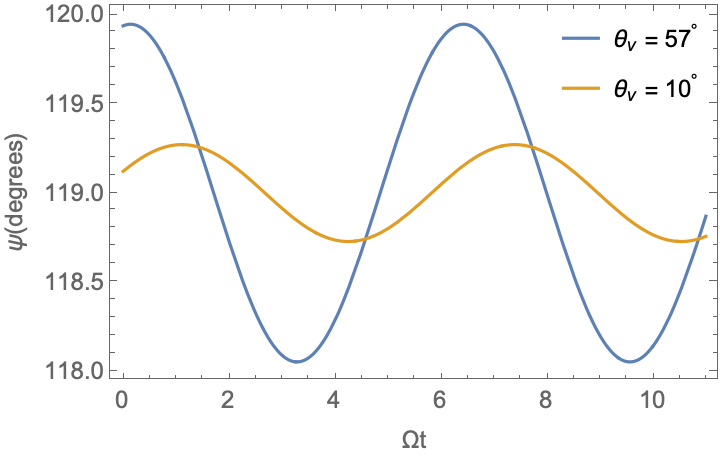}
    \caption{Oscillating linear polarization angle of the crab pulsar (blue) due to the axion field sourced by the pulsar's magnetic field $g_{a \gamma \gamma} = 1.5 \times 10^{-10}~{\rm GeV}^{-1}$. The yellow curve shows the linear polarization angle of an identical pulsar with viewing angle $\theta_v = 10^{\circ}$.}
    \label{psi}
\end{figure}

For the Crab pulsar, $\alpha \sim 45^\circ - 75^\circ$ and $\theta_v \sim 57^\circ$ \cite{Bogovalov:2018had}. The observed linear polarization angle of the unpulsed emission is $\psi = 119^{\circ}$ \cite{Slowikowska:2009rm}. The resulting observed oscillating linear polarization angle for $g_{a \gamma \gamma} = 1.5 \times 10^{-10}~{\rm GeV}$ derived from Eq.~\eqref{totalStokes} is shown in FIG.~\eqref{psi}. For comparison, we also show the observed oscillating linear polarization angle for an identical pulsar viewed at an angle $\theta_v = 10^{\circ}$ to its rotation axis. The linear polarization degree would also oscillate with frequency $\Omega$, but the amplitude of these oscillations is negligible compared to the oscillations in $\psi$.

Based on the fact that the unpulsed's component's linear polarization angle is constant to within $\sim 1^{\circ}$ \cite{Slowikowska:2009rm}, we can constrain the axion-photon coupling to

\begin{equation}
g_{a\gamma\gamma} \lesssim 1.5\times10^{-10}~\mathrm{GeV^{-1}},
\end{equation}
where the sensitivity applies to axion masses $m_a << \Omega = 1.2\times10^{-13}~\mathrm{eV}$. The time dependence of the polarization angle for this value of $g_{a \gamma \gamma}$ is shown in FIG.~\eqref{psi}. The exact amplitude depends on the distribution of polarization and intensity of the unpulsed emission across the surface of the pulsar. However, the key features of the axion-induced birefringence signal are independent of the pulsar emission structure. Firstly, the induced birefringent rotation angle is independent of the photon frequency. Thus, the axion-induced birefringent effect can be disentangled from the plasma-induced dispersion measure effect, which varies as $1/\omega^2$. The axion-induced birefringence is periodically modulated in time and phase locked to the stellar rotation frequency $\Omega$. Furthermore, as discussed above, a generic feature of birefringence induced by the pulsar sourced axion field is that the effect is greater for larger pulsar viewing angles, as well as for larger magnetic fields. This effect does not rely on the axion constituting the DM.

The constraint on $g_{a \gamma \gamma}$ derived from the Crab pulsar is weaker than existing bounds on $g_{a \gamma \gamma}$. Future polarization measurements with improved precision, as well as observations of millisecond pulsars and compact stars with larger magnetic fields, such as magnetars, can further strengthen these bounds.

The leading astrophysical bounds on $g_{a \gamma \gamma}$ for ultra-light axions not comprising DM arise from non-observation of spectral modulations in X-ray point sources shining through the magnetic fields of galaxy clusters, yielding $g_{a \gamma \gamma} \lesssim 6.3\times10^{-13}~\mathrm{GeV^{-1}}$ for $m_a \lesssim 10^{-12}~\mathrm{eV}$ \cite{Reynes:2021bpe}. For polarimetry observations of the Crab pulsar to become competitive with these bounds using the method presented above, the polarisation angle measurement precision would need to improve by a factor of $10^4$. This is of course rather unfeasible. The Crab pulsar was chosen for this proof of principle study due to the availability of time resolved polarimetry measurements and estimates of key parameters. However, millisecond {\it magnetars}, hosting much higher magnetic fields, are the ideal laboaratories for this effect. For example, observations of magnetars with magnetic fields $B \sim 10^{16} \, {\rm G}$ and rotational periods $P \sim 1 \, {\rm s}$ are discussed in \cite{Rowlinson:2013ue}. If the time resolved polarisation angle of these magnetars could be measured to a precision of $\sim 1^{\circ}$, the bound on $g_{a \gamma \gamma}$ could be improved to $g_{a \gamma \gamma} \lesssim 10^{-14}~\mathrm{GeV^{-1}}$.

\section{Conclusions and discussions}\label{sec5}

Neutron stars provide powerful laboratories for probing fundamental physics in extreme environments. The increasing sensitivity of modern telescopes in the gamma-ray, X-ray, and optical bands opens new opportunities to test beyond SM scenarios that cannot be probed in terrestrial experiments. In this work, we have used the Crab pulsar as a testbed to investigate photon interactions with ultralight axions through the axion-photon coupling.

We computed the axion-induced rotation of the polarization angle of pulsar emission and use time-resolved polarimetric observations of the Crab pulsar to place bounds on the axion-photon coupling. This birefringence effect does not rely on axions comprising the DM and shows distinctive features such as frequency independence and dependence on the viewing angle of the pulsar. Future high-precision polarimetric and spectroscopic observations of pulsars and magnetars therefore have the potential to improve these constraints by several orders of magnitude. Such observations would also motivate a more sophisticated analysis, for example, using a Fourier space analysis to search for the characteristic oscillations of the polarisation angle, or searching for the expected correlations between the magnitude of this effect and the pulsar viewing angle. In this way, NSs may be established as sensitive probes of ultralight axions and other light new degrees of freedom. 

\section*{Acknowledgments} 
T.K.P would like to thank Abdus Salam International Centre for Theoretical Physics (ICTP) for its kind hospitality where part of the work has been done. The authors would also like to thank Alexei Y Smirnov for useful comments and suggestions. This article is based on the work from COST Actions COSMIC WISPers CA21106 and BridgeQG CA23130, supported by COST (European Cooperation in Science and Technology) The authors are supported by STFC under ST/X000745/1 and ST/X003167/1. For the purpose of open access, the authors has applied a creative commons attribution (CC BY) licence to any author accepted manuscript version arising.

\section*{Data Availability Statement}
This article has no associated data.

\appendix
\section{Time-independent axion field profile}\label{appendix}
To calculate the time-independent massless axion field profile, we write the axion equation of motion 
\begin{equation}
\nabla^2a(r,\theta)=-g_{a\gamma\gamma}\mathbf{E}\cdot \mathbf{B},   
\label{app1}
\end{equation}
where the static axion-induced source-charge density for $r>R$ is obtained as
\begin{equation}
J(r,\theta)=g_{a\gamma\gamma}\mathbf{E}\cdot \mathbf{B}=-g_{a\gamma\gamma}\frac{B^2_0\Omega R^8}{r^7}\cos^3\theta=-\frac{\mathcal{C}}{r^7}\cos^3\theta,
\label{app2}
\end{equation}
where $\mathcal{C}=g_{a\gamma\gamma}B^2_0\Omega R^8$, and Eq.~\eqref{eq:3} is used for $\mathbf{B}$ and $\mathbf{E}$.

We use the boundary conditions that the axion field is regular at $r\to 0$ and $a\to 0$ as $r\to \inf$. Additionally, we consider that the axion field profile and its derivative are continuous at the surface $r=R$. The axion-induced source charge density exists only outside the star and $J(r,\theta)=0$ for $r<R$. 

In deriving the axion field profile, we start with the Green's function
\begin{equation}
G(\mathbf{x},\mathbf{x}^\prime)=-\frac{1}{4\pi|\mathbf{x}-\mathbf{x}^\prime|},~~~\nabla^2G(\mathbf{x},\mathbf{x}^\prime)=\delta^3(\mathbf{x}-\mathbf{x}^\prime),
\label{app3}
\end{equation}
which results the axion field profile 
\begin{equation}
a(\mathbf{x})=\int d^3x^\prime G(\mathbf{x},\mathbf{x}^\prime)J(\mathbf{x}^\prime)=-\frac{1}{4\pi}\int d^3 x^\prime \frac{J(\mathbf{x}^\prime)}{|\mathbf{x}-\mathbf{x}^\prime|}.   
\label{app4}
\end{equation}
The above axion field can be expressed in terms of Legendre polynomials by using the expansion
\begin{equation}
\frac{1}{|\mathbf{x}-\mathbf{x}^\prime|}=\sum ^\inf_{l=0}\frac{r^l_<}{r^{l+1}_>}P_l(\cos\gamma),
\label{app5}
\end{equation}
where $r_<\equiv\mathrm{min}(r,r^\prime)$, $r_>\equiv\mathrm{max}(r,r^\prime)$, and $\gamma$ is the angle between $\mathbf{x}$ and $\mathbf{x}^\prime$ with $\cos\gamma=\cos\theta\cos\theta^\prime+\sin\theta\sin\theta^\prime \cos(\varphi-\varphi^\prime)$. 

Therefore, we can write the axion field in terms of mode expansion as
\begin{equation}
a(r,\theta)=\sum ^\inf_{l=0}a_l(r)P_l(\cos\theta),    
\label{app6}
\end{equation}
with radial modes
\begin{equation}
\begin{split}
a_l(r)=-\frac{1}{2l+1}\Big[\frac{1}{r^{l+1}}\int ^r_R dr ^\prime {r^\prime}^{l+2}J_l(r^\prime)+\\
r^l\int ^\inf_r dr^\prime {r^\prime}^{1-l}J_l(r^\prime) \Big],
\end{split}
\label{app7}
\end{equation}
where the Legendre coefficients of the source are 
\begin{equation}
J_l(r)=\frac{2l+1}{2}\int ^{+1}_{-1}d\mu P_l(\mu)J(r,\mu), ~\mu=\cos\theta.
\label{app8}
\end{equation}
Using the identity $\mu^3=\cos^3\theta=(3/5)P_1(\mu)+(2/5)P_3(\mu)$, we write Eq.~\ref{app2} as
\begin{equation}
J(r,\mu)=-\frac{\mathcal{C}}{r^7}\mu^3=-\frac{\mathcal{C}}{r^7}\Big[\frac{3}{5}P_1(\mu)+\frac{2}{5}P_2(\mu)\Big].  
\label{app9}
\end{equation}
Therefore, $l=1,3$ modes will contribute to the axion field profile and $J_l(r)=0$ for $l\neq 1,3$. However, the dominant contribution would come from $l=1$ mode and we obtain the radial mode from Eq.~\ref{app7} to the leading order as
\begin{equation}
a_1(r)\simeq \frac{\mathcal{C}}{15R^3 r^2}.   
\end{equation}
Thus, using  Eq.~\ref{app6}, we obtain the axion field profile outside the star to the leading order as 
\begin{equation}
a(r,\theta)\simeq \frac{g_{a\gamma\gamma}B^2_0R^5\Omega}{15r^2}\cos\theta,    
\end{equation}
which shows a dipolar behavior.

\bibliographystyle{utphys}
\bibliography{reference}

\providecommand{\href}[2]{#2}\begingroup\raggedright\begin{thebibliography}{10}

\bibitem{Peccei:1977hh}
R.~D. Peccei and H.~R. Quinn, ``{CP Conservation in the Presence of
  Instantons},'' \href{http://dx.doi.org/10.1103/PhysRevLett.38.1440}{{\em
  Phys. Rev. Lett.} {\bfseries 38} (1977) 1440--1443}.

\bibitem{Weinberg:1977ma}
S.~Weinberg, ``{A New Light Boson?},''
  \href{http://dx.doi.org/10.1103/PhysRevLett.40.223}{{\em Phys. Rev. Lett.}
  {\bfseries 40} (1978) 223--226}.

\bibitem{Peccei:1977ur}
R.~D. Peccei and H.~R. Quinn, ``{Constraints Imposed by CP Conservation in the
  Presence of Instantons},''
  \href{http://dx.doi.org/10.1103/PhysRevD.16.1791}{{\em Phys. Rev. D}
  {\bfseries 16} (1977) 1791--1797}.

\bibitem{Wilczek:1977pj}
F.~Wilczek, ``{Problem of Strong $P$ and $T$ Invariance in the Presence of
  Instantons},'' \href{http://dx.doi.org/10.1103/PhysRevLett.40.279}{{\em Phys.
  Rev. Lett.} {\bfseries 40} (1978) 279--282}.

\bibitem{Adler:1969gk}
S.~L. Adler, ``{Axial vector vertex in spinor electrodynamics},''
  \href{http://dx.doi.org/10.1103/PhysRev.177.2426}{{\em Phys. Rev.} {\bfseries
  177} (1969) 2426--2438}.

\bibitem{Bell:1969ts}
J.~S. Bell and R.~Jackiw, ``{A PCAC puzzle: $\pi^0 \to \gamma \gamma$ in the
  $\sigma$ model},'' \href{http://dx.doi.org/10.1007/BF02823296}{{\em Nuovo
  Cim. A} {\bfseries 60} (1969) 47--61}.

\bibitem{Peccei:2006as}
R.~D. Peccei, ``{The Strong CP problem and axions},''
  \href{http://dx.doi.org/10.1007/978-3-540-73518-2_1}{{\em Lect. Notes Phys.}
  {\bfseries 741} (2008) 3--17},
  \href{http://arxiv.org/abs/hep-ph/0607268}{{\ttfamily arXiv:hep-ph/0607268}}.

\bibitem{Baker:2006ts}
C.~A. Baker {\em et~al.}, ``{An Improved experimental limit on the electric
  dipole moment of the neutron},''
  \href{http://dx.doi.org/10.1103/PhysRevLett.97.131801}{{\em Phys. Rev. Lett.}
  {\bfseries 97} (2006) 131801},
  \href{http://arxiv.org/abs/hep-ex/0602020}{{\ttfamily arXiv:hep-ex/0602020}}.

\bibitem{Kim:2008hd}
J.~E. Kim and G.~Carosi, ``{Axions and the Strong CP Problem},''
  \href{http://dx.doi.org/10.1103/RevModPhys.82.557}{{\em Rev. Mod. Phys.}
  {\bfseries 82} (2010) 557--602},
  \href{http://arxiv.org/abs/0807.3125}{{\ttfamily arXiv:0807.3125 [hep-ph]}}.
  [Erratum: Rev.Mod.Phys. 91, 049902 (2019)].

\bibitem{GrillidiCortona:2015jxo}
G.~Grilli~di Cortona, E.~Hardy, J.~Pardo~Vega, and G.~Villadoro, ``{The QCD
  axion, precisely},'' \href{http://dx.doi.org/10.1007/JHEP01(2016)034}{{\em
  JHEP} {\bfseries 01} (2016) 034},
  \href{http://arxiv.org/abs/1511.02867}{{\ttfamily arXiv:1511.02867
  [hep-ph]}}.

\bibitem{Preskill:1982cy}
J.~Preskill, M.~B. Wise, and F.~Wilczek, ``{Cosmology of the Invisible
  Axion},'' \href{http://dx.doi.org/10.1016/0370-2693(83)90637-8}{{\em Phys.
  Lett. B} {\bfseries 120} (1983) 127--132}.

\bibitem{Abbott:1982af}
L.~F. Abbott and P.~Sikivie, ``{A Cosmological Bound on the Invisible Axion},''
  \href{http://dx.doi.org/10.1016/0370-2693(83)90638-X}{{\em Phys. Lett. B}
  {\bfseries 120} (1983) 133--136}.

\bibitem{Dine:1982ah}
M.~Dine and W.~Fischler, ``{The Not So Harmless Axion},''
  \href{http://dx.doi.org/10.1016/0370-2693(83)90639-1}{{\em Phys. Lett. B}
  {\bfseries 120} (1983) 137--141}.

\bibitem{Svrcek:2006yi}
P.~Svrcek and E.~Witten, ``{Axions In String Theory},''
  \href{http://dx.doi.org/10.1088/1126-6708/2006/06/051}{{\em JHEP} {\bfseries
  06} (2006) 051}, \href{http://arxiv.org/abs/hep-th/0605206}{{\ttfamily
  arXiv:hep-th/0605206}}.

\bibitem{Marsh:2015xka}
D.~J.~E. Marsh, ``{Axion Cosmology},''
  \href{http://dx.doi.org/10.1016/j.physrep.2016.06.005}{{\em Phys. Rept.}
  {\bfseries 643} (2016) 1--79},
  \href{http://arxiv.org/abs/1510.07633}{{\ttfamily arXiv:1510.07633
  [astro-ph.CO]}}.

\bibitem{Arvanitaki:2009fg}
A.~Arvanitaki, S.~Dimopoulos, S.~Dubovsky, N.~Kaloper, and J.~March-Russell,
  ``{String Axiverse},''
  \href{http://dx.doi.org/10.1103/PhysRevD.81.123530}{{\em Phys. Rev. D}
  {\bfseries 81} (2010) 123530},
  \href{http://arxiv.org/abs/0905.4720}{{\ttfamily arXiv:0905.4720 [hep-th]}}.

\bibitem{Hu:2000ke}
W.~Hu, R.~Barkana, and A.~Gruzinov, ``{Cold and fuzzy dark matter},''
  \href{http://dx.doi.org/10.1103/PhysRevLett.85.1158}{{\em Phys. Rev. Lett.}
  {\bfseries 85} (2000) 1158--1161},
  \href{http://arxiv.org/abs/astro-ph/0003365}{{\ttfamily
  arXiv:astro-ph/0003365}}.

\bibitem{Duffy:2009ig}
L.~D. Duffy and K.~van Bibber, ``{Axions as Dark Matter Particles},''
  \href{http://dx.doi.org/10.1088/1367-2630/11/10/105008}{{\em New J. Phys.}
  {\bfseries 11} (2009) 105008},
  \href{http://arxiv.org/abs/0904.3346}{{\ttfamily arXiv:0904.3346 [hep-ph]}}.

\bibitem{Hui:2016ltb}
L.~Hui, J.~P. Ostriker, S.~Tremaine, and E.~Witten, ``{Ultralight scalars as
  cosmological dark matter},''
  \href{http://dx.doi.org/10.1103/PhysRevD.95.043541}{{\em Phys. Rev. D}
  {\bfseries 95} no.~4, (2017) 043541},
  \href{http://arxiv.org/abs/1610.08297}{{\ttfamily arXiv:1610.08297
  [astro-ph.CO]}}.

\bibitem{Chadha-Day:2021szb}
F.~Chadha-Day, J.~Ellis, and D.~J.~E. Marsh, ``{Axion dark matter: What is it
  and why now?},'' \href{http://dx.doi.org/10.1126/sciadv.abj3618}{{\em Sci.
  Adv.} {\bfseries 8} no.~8, (2022) abj3618},
  \href{http://arxiv.org/abs/2105.01406}{{\ttfamily arXiv:2105.01406
  [hep-ph]}}.

\bibitem{OHare:2024nmr}
C.~A.~J. O'Hare, ``{Cosmology of axion dark matter},''
  \href{http://dx.doi.org/10.22323/1.454.0040}{{\em PoS} {\bfseries
  COSMICWISPers} (2024) 040}, \href{http://arxiv.org/abs/2403.17697}{{\ttfamily
  arXiv:2403.17697 [hep-ph]}}.

\bibitem{Profumo:2017hqp}
S.~Profumo, \href{http://dx.doi.org/10.1142/q0001}{{\em {An Introduction to
  Particle Dark Matter}}}.
\newblock World Scientific, 2017.

\bibitem{Moody:1984ba}
J.~E. Moody and F.~Wilczek, ``{NEW MACROSCOPIC FORCES?},''
  \href{http://dx.doi.org/10.1103/PhysRevD.30.130}{{\em Phys. Rev. D}
  {\bfseries 30} (1984) 130}.

\bibitem{AxionLimits}
C.~O'Hare, ``cajohare/axionlimits: Axionlimits.''
  \url{https://cajohare.github.io/AxionLimits/}, July, 2020.

\bibitem{Moran:2013cla}
P.~Moran, A.~Shearer, R.~Mignani, A.~S{\l}owikowska, A.~De~Luca,
  C.~Gouiff{\`e}s, and P.~Laurent, ``{Optical Polarimetry of the Inner Crab
  Nebula and Pulsar},'' \href{http://dx.doi.org/10.1093/mnras/stt931}{{\em Mon.
  Not. Roy. Astron. Soc.} {\bfseries 433} (2013) 2564},
  \href{http://arxiv.org/abs/1305.6824}{{\ttfamily arXiv:1305.6824
  [astro-ph.HE]}}.

\bibitem{weisskopf22}
M.~C. {Weisskopf}, P.~{Soffitta}, L.~{Baldini}, B.~D. {Ramsey}, S.~L. {O'Dell},
  and R.~et~al., ``{The Imaging X-Ray Polarimetry Explorer (IXPE):
  Pre-Launch},'' \href{http://dx.doi.org/10.1117/1.JATIS.8.2.026002}{{\em
  Journal of Astronomical Telescopes, Instruments, and Systems} {\bfseries 8}
  no.~2, (Apr., 2022) 026002},
  \href{http://arxiv.org/abs/2112.01269}{{\ttfamily arXiv:2112.01269
  [astro-ph.IM]}}.

\bibitem{Gonzalez-Caniulef:2024qsf}
D.~Gonz{\'a}lez-Caniulef, J.~Heyl, S.~Fabiani, P.~Soffitta, E.~Costa,
  N.~Bucciantini, D.~Kirmizibayrak, and F.~Xie, ``{Crab pulsar: IXPE
  observations reveal unified polarization properties in the optical and soft
  X-ray bands},'' \href{http://dx.doi.org/10.1051/0004-6361/202451815}{{\em
  Astron. Astrophys.} {\bfseries 693} (2025) A152},
  \href{http://arxiv.org/abs/2408.03245}{{\ttfamily arXiv:2408.03245
  [astro-ph.HE]}}.

\bibitem{Dean:2008zz}
A.~J. Dean, D.~J. Clark, J.~B. Stephen, V.~A. McBride, L.~Bassani, A.~Bazzano,
  A.~J. Bird, A.~B. Hill, S.~E. Shaw, and P.~Ubertini, ``{Polarized gamma-ray
  emission from the Crab},''
  \href{http://dx.doi.org/10.1126/science.1149056}{{\em Science} {\bfseries
  321} (2008) 1183--1185}.

\bibitem{POLARBEAR:2024vel}
{\bfseries POLARBEAR} Collaboration, S.~Adachi {\em et~al.}, ``{Exploration of
  the polarization angle variability of the Crab Nebula with POLARBEAR and its
  application to the search for axionlike particles},''
  \href{http://dx.doi.org/10.1103/PhysRevD.110.063013}{{\em Phys. Rev. D}
  {\bfseries 110} no.~6, (2024) 063013},
  \href{http://arxiv.org/abs/2403.02096}{{\ttfamily arXiv:2403.02096
  [astro-ph.CO]}}.

\bibitem{Liu:2019brz}
T.~Liu, G.~Smoot, and Y.~Zhao, ``{Detecting axionlike dark matter with linearly
  polarized pulsar light},''
  \href{http://dx.doi.org/10.1103/PhysRevD.101.063012}{{\em Phys. Rev. D}
  {\bfseries 101} no.~6, (2020) 063012},
  \href{http://arxiv.org/abs/1901.10981}{{\ttfamily arXiv:1901.10981
  [astro-ph.CO]}}.

\bibitem{SPT-3G:2022ods}
{\bfseries SPT-3G} Collaboration, K.~R. Ferguson {\em et~al.}, ``{Searching for
  axionlike time-dependent cosmic birefringence with data from SPT-3G},''
  \href{http://dx.doi.org/10.1103/PhysRevD.106.042011}{{\em Phys. Rev. D}
  {\bfseries 106} no.~4, (2022) 042011},
  \href{http://arxiv.org/abs/2203.16567}{{\ttfamily arXiv:2203.16567
  [astro-ph.CO]}}.

\bibitem{Fan:2025ixw}
J.~Fan, L.~Li, and C.~Sun, ``{Pulse and Polarization Structures in
  Axion-Converted X-Rays from Pulsars},''
  \href{http://dx.doi.org/10.1103/66m2-6w2p}{{\em Phys. Rev. Lett.} {\bfseries
  135} no.~23, (2025) 231801},
  \href{http://arxiv.org/abs/2501.12440}{{\ttfamily arXiv:2501.12440
  [hep-ph]}}.

\bibitem{deutsch}
A.~J. {Deutsch}, ``{The electromagnetic field of an idealized star in rigid
  rotation in vacuo},'' {\em Annales d'Astrophysique} {\bfseries 18} (Jan.,
  1955) 1.

\bibitem{Khelashvili:2024sup}
M.~Khelashvili, M.~Lisanti, A.~Prabhu, and B.~R. Safdi, ``{Axion
  pulsarscope},'' \href{http://dx.doi.org/10.1103/PhysRevD.111.083027}{{\em
  Phys. Rev. D} {\bfseries 111} no.~8, (2025) 083027},
  \href{http://arxiv.org/abs/2402.17820}{{\ttfamily arXiv:2402.17820
  [hep-ph]}}.

\bibitem{Garbrecht:2018akc}
B.~Garbrecht and J.~I. McDonald, ``{Axion configurations around pulsars},''
  \href{http://dx.doi.org/10.1088/1475-7516/2018/07/044}{{\em JCAP} {\bfseries
  07} (2018) 044}, \href{http://arxiv.org/abs/1804.04224}{{\ttfamily
  arXiv:1804.04224 [astro-ph.CO]}}.

\bibitem{Blas:2019qqp}
D.~Blas, A.~Caputo, M.~M. Ivanov, and L.~Sberna, ``{No chiral light bending by
  clumps of axion-like particles},''
  \href{http://dx.doi.org/10.1016/j.dark.2019.100428}{{\em Phys. Dark Univ.}
  {\bfseries 27} (2020) 100428},
  \href{http://arxiv.org/abs/1910.06128}{{\ttfamily arXiv:1910.06128
  [hep-ph]}}.

\bibitem{PhysRev.126.1899}
S.~Weinberg, ``Eikonal method in magnetohydrodynamics,''
  \href{http://dx.doi.org/10.1103/PhysRev.126.1899}{{\em Phys. Rev.} {\bfseries
  126} (Jun, 1962) 1899--1909}.
  \url{https://link.aps.org/doi/10.1103/PhysRev.126.1899}.

\bibitem{Lyne:2014qqa}
A.~Lyne, C.~Jordan, F.~Graham-Smith, C.~Espinoza, B.~Stappers, and
  P.~Weltrvrede, ``{45 years of rotation of the Crab pulsar},''
  \href{http://dx.doi.org/10.1093/mnras/stu2118}{{\em Mon. Not. Roy. Astron.
  Soc.} {\bfseries 446} (2015) 857--864},
  \href{http://arxiv.org/abs/1410.0886}{{\ttfamily arXiv:1410.0886
  [astro-ph.HE]}}.

\bibitem{Slowikowska:2009rm}
A.~Slowikowska, G.~Kanbach, M.~Kramer, and A.~Stefanescu, ``{Optical
  polarisation of the Crab pulsar: Precision measurements and comparison to the
  radio emission},''
  \href{http://dx.doi.org/10.1111/j.1365-2966.2009.14935.x}{{\em Mon. Not. Roy.
  Astron. Soc.} {\bfseries 397} (2009) 103},
  \href{http://arxiv.org/abs/0901.4559}{{\ttfamily arXiv:0901.4559
  [astro-ph.SR]}}.

\bibitem{Bogovalov:2018had}
S.~V. Bogovalov, I.~Contopoulos, A.~Prosekin, I.~Tronin, and F.~A. Aharonian,
  ``{Magnetic absorption of VHE photons in the magnetosphere of the Crab
  pulsar},'' \href{http://dx.doi.org/10.1093/mnras/sty455}{{\em Mon. Not. Roy.
  Astron. Soc.} {\bfseries 476} no.~3, (2018) 4213--4223},
  \href{http://arxiv.org/abs/1902.01191}{{\ttfamily arXiv:1902.01191
  [astro-ph.HE]}}.

\bibitem{Reynes:2021bpe}
J.~S. Reyn{\'e}s, J.~H. Matthews, C.~S. Reynolds, H.~R. Russell, R.~N. Smith,
  and M.~C.~D. Marsh, ``{New constraints on light axion-like particles using
  Chandra transmission grating spectroscopy of the powerful cluster-hosted
  quasar H1821+643},'' \href{http://dx.doi.org/10.1093/mnras/stab3464}{{\em
  Mon. Not. Roy. Astron. Soc.} {\bfseries 510} no.~1, (2021) 1264--1277},
  \href{http://arxiv.org/abs/2109.03261}{{\ttfamily arXiv:2109.03261
  [astro-ph.HE]}}.

\bibitem{Rowlinson:2013ue}
A.~Rowlinson, P.~T. O'Brien, B.~D. Metzger, N.~R. Tanvir, and A.~J. Levan,
  ``{Signatures of magnetar central engines in short GRB lightcurves},''
  \href{http://dx.doi.org/10.1093/mnras/sts683}{{\em Mon. Not. Roy. Astron.
  Soc.} {\bfseries 430} (2013) 1061},
  \href{http://arxiv.org/abs/1301.0629}{{\ttfamily arXiv:1301.0629
  [astro-ph.HE]}}.

\end{thebibliography}\endgroup
\end{document}